\documentclass[doublecol,figures]{epl2} 

\usepackage{amsmath}
\usepackage{color}

\newcommand{\rmd}{\mathrm{d}}
\newcommand{\rme}{\mathrm{e}}
\newcommand{\sub}[1]{_\mathrm{#1}}

\title{Strong anomalous diffusion of the phase of a chaotic pendulum}
\shorttitle{Strong anomalous diffusion of the phase of a chaotic pendulum} 

\author{Francesco Cagnetta\inst{1} \and Giuseppe Gonnella\inst{1} \and Alessandro Mossa\inst{1} \and Stefano Ruffo\inst{2}}
\shortauthor{F. Cagnetta, G. Gonnella, A. Mossa and S. Ruffo}

\institute{                    
  \inst{1} Dipartimento di Fisica, Universit\`a di Bari and INFN, Sezione di Bari - Via Amendola 173, 70126 Bari, Italy\\
  \inst{2} Dipartimento di Fisica e Astronomia and CSDC, Universit\`a di Firenze, CNISM and INFN - Via Sansone 1, 50019 Sesto Fiorentino, Italy 
}
\pacs{05.45.-a}{Nonlinear dynamics and chaos}
\pacs{05.40.Fb}{Random walks and L{\'e}vy flights}
\pacs{05.60.-k}{Transport processes}

\abstract{
In this letter we consider the phase diffusion of a harmonically driven undamped pendulum and show that it is anomalous in the strong sense. The role played by the fractal properties of the phase space is highlighted, providing an illustration of the link between deterministic chaos and anomalous transport. Finally, we build a stochastic model which reproduces most properties of the original Hamiltonian system by alternating ballistic flights and random diffusion.
}

\begin{document}

\maketitle

Anomalous transport~\cite{Klages:2008} is a phenomenon relevant to a wide range of complex systems and has recently attracted the attention of researchers with a mathematical, physical, chemical, biological and socio-economical background. The list of diffusion processes where violations of the hypotheses of the central limit theorem lead to departures from the normal long-time asymptotic behaviour $\langle x^2(t)\rangle\sim t$, where $x(t)$ is a generic dynamical variable, is huge and ever-growing due to an intense theoretical and experimental research effort~\cite{Bouchaud:1990}. Here we focus on strong anomalous diffusion~\cite{Castiglione:1999}, defined by the property
\begin{equation} \label{eq:str}
	\langle |x(t)|^q \rangle \sim t^{q\nu(q)}\,,
\end{equation}
where the function $\nu(q)$ is not constant. Such behaviour has been detected in numerical studies over a variety of systems: without any claim of completeness, we can cite intermittent 1D maps~\cite{Pikovsky:1991}, running sandpile models~\cite{Carreras:1999}, infinite horizon~\cite{Armstead:2003} and polygonal~\cite{Sanders:2006}  billiards, cold atoms in optical lattices~\cite{Dechant:2012} and 1D inhomogeneous materials~\cite{Bernabo:2014}. The emergence of strong anomalous diffusion has been analytically investigated in the context of both stochastic~\cite{Andersen:2000,Rebenshtok:2014} and Hamiltonian~\cite{Artuso:2003} models. The first (and, at the moment, only) experimental measure of strong anomalous diffusion has been very recently obtained by tracking polymeric particles in living cells~\cite{Gal:2010}. In this letter we will  investigate the diffusion properties of the phase of a harmonically driven pendulum. This system is  representative of the features of generic Hamiltonian continuous-time dynamics but at the same time it is simple enough to be considered a paradigmatic  model for anomalous transport.
 
\begin{figure*}
\begin{center}
\includegraphics{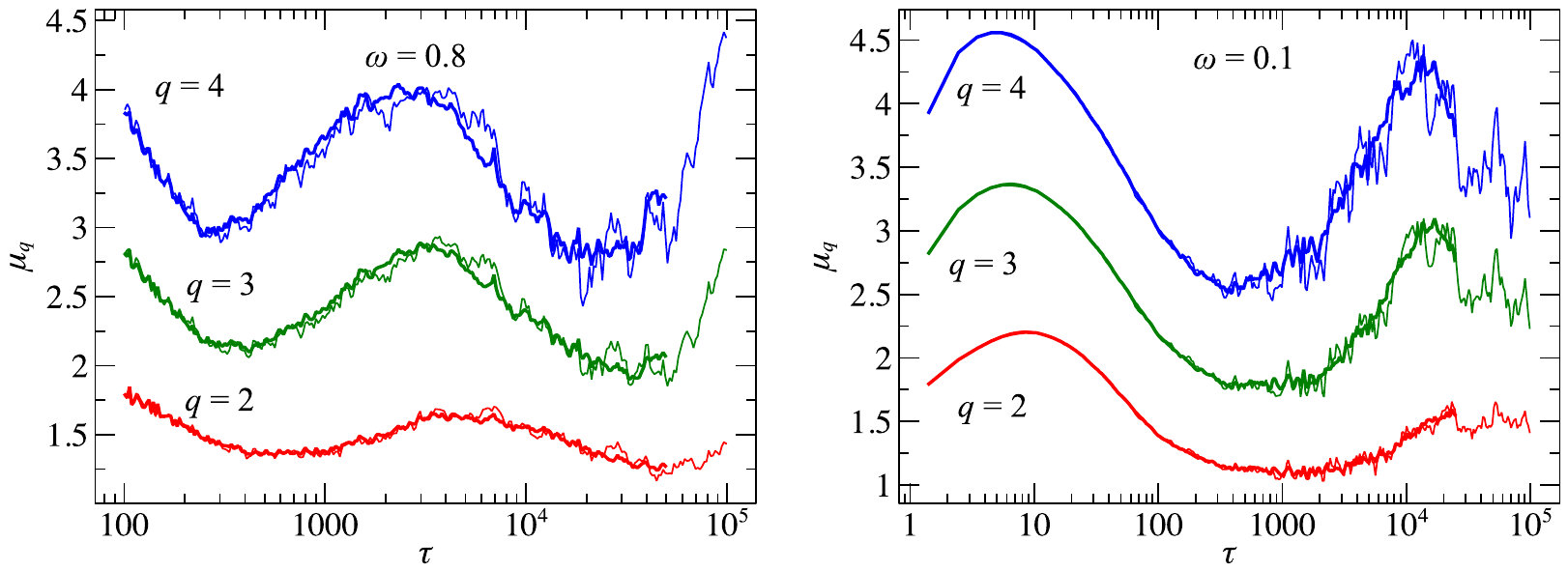}
\caption{Function $\mu_q(\tau)$ as defined by eq.~(\ref{eq:qnuq}) for different values of $q$. In order to show that the log-periodic oscillations are not an artifact due to poor sampling, we have computed $10^5$ additional (shorter) traces, which contribute to the thick lines.}
\label{fig:wav}
\end{center}
\end{figure*}

\section{Chaotic pendulum}
The pendulum is the most familiar nonlinear system~\cite{Baker:2005} and its study goes back to the very first appearance of modern (as opposed to ancient) physics. When subject to a suitable periodic driving force, the pendulum displays chaotic behaviour~\cite{Gitterman:2010}; one can then investigate its phase diffusion properties~\cite{Blackburn:1996} by choosing many initial conditions in a tiny region of the phase space and observing how the deterministic trajectories spread out. 
The periodically driven undamped pendulum equation we consider here
\begin{equation} \label{eq:pend}
	\ddot{\theta}+\sin\theta = \gamma\sin(\omega t)
\end{equation} 
has been studied previously by the authors of refs.~\cite{Harish:2002,Harish:2003,Harish:2006,Sakthivel:2011}, who report instances of both regular (in the case $\gamma=1.2$, $\omega=0.1$) and anomalous (for $\gamma=1.2$, $\omega=0.8$) diffusion. Actually, we partially correct here those statements, showing explicitly that the diffusion is strongly anomalous in both cases.  

There are good reasons to believe that strong anomalous diffusion is rather the rule than the exception in Hamiltonian systems  characterised by a mixed phase space, where regular islands are surrounded by the chaotic sea~\cite{Zaslavsky:2002,Zaslavsky:2007, Leoncini:2008}. A chaotic trajectory cannot enter an island, which is formed by KAM-tori, yet it can temporarily behave like a regular motion due to the presence of \emph{cantori}~\cite{MacKay:1984} all around the island. Cantori are fractal objects that look like closed curves with an infinite number of gaps~\cite{Makarov:2010}; they act on trajectories as quasi-traps, i.e., they cause the velocity to remain constant during sizeable periods of time, whereas a purely chaotic motion would lead to nearly uncorrelated displacements. 
 The kinetics is then similar to that of a L\'evy walk~\cite{Klafter:1994,Denisov:2002}, a process where the system randomly select a velocity and maintains it for a random time before renewing the choice. Such process is known to give rise to strong anomalous diffusion~\cite{Andersen:2000}. The previous quite general argument suggests that there is nothing special in our choice of the parameters $(\gamma, \omega)$ and that the same conclusions would be obtained by inspecting any other value compatible with chaotic behaviour.

In the present work we numerically integrate eq.~(\ref{eq:pend}) using a Runge--Kutta 4th order algorithm~\cite{Press:2007}. The adequacy of the chosen integration step ($\delta t=0.005$ for $\omega=0.1$ and $\delta t = 0.001$ for $\omega = 0.8$) was confirmed by checking that no statistical property of the trajectories is significantly altered when the integration step is halved. For each value of $\omega$, about $5\times10^4$ initial conditions $(\theta_0,\dot{\theta}_0)$ are randomly chosen in the region $-0.05<\theta_0<0.05$, $-0.05<\dot{\theta}_0<0.05$. The time $t$ appearing in eq.~(\ref{eq:pend}) is measured in units of the proper frequency of the pendulum, but in order to compare behaviours for different values of $\omega$ it is useful to present results in terms of the normalised time $\tau$, measured in cycles of the external force: $\tau = \omega t/(2\pi)$. We follow the evolution of our sample trajectories along $10^5$ such cycles.

\begin{figure}
\onefigure{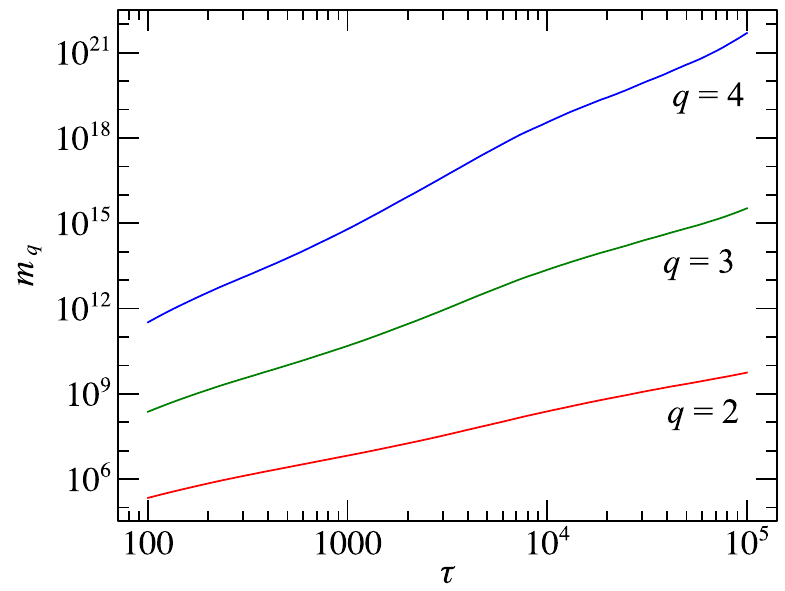}
\caption{Statistical momenta $m_q$ as a function of the normalised time $\tau$, for $\omega=0.8$. (The case $\omega=0.1$ yields very similar results.)}
\label{fig:mom}
\end{figure}

\section{Phase diffusion properties}
There are a few standard ways to characterise anomalous diffusion~\cite{Schmiedeberg:2009}. One is to evaluate the asymptotic exponent $q\nu(q)$ of fractional moments $m_q(\tau)$:
\begin{equation} \label{eq:qnu}
	m_q(\tau) \equiv \langle|\theta(\tau)|^q\rangle\sim \tau^{q\nu(q)} \quad \text{for } \tau\to\infty \,. 
\end{equation}
The study of $m_q(\tau)$ over several orders of magnitude in time reveals small but not negligible departures from a straight power law  (see fig.~\ref{fig:mom}), which are more clearly visualised if we consider the quantity 
\begin{equation} \label{eq:qnuq}
	\mu_q(\tau) \equiv \tau \frac{\rmd}{\rmd \tau} \ln m_q(\tau) \,,
\end{equation}	
showed in fig.~\ref{fig:wav}. Similar log-periodic oscillations have been related to the presence of self-similar structures in the phase space~\cite{Meiss:1986} and occur frequently in the literature about diffusion on fractals~\cite{Grabner:1997,Padilla:2009,Derrida:2014}. For our purposes here, the existence of these oscillations implies that the asymptotic exponent $q\nu(q)$ markedly depends on the time interval over which it is estimated. 

\begin{figure*}
\begin{center}
\includegraphics{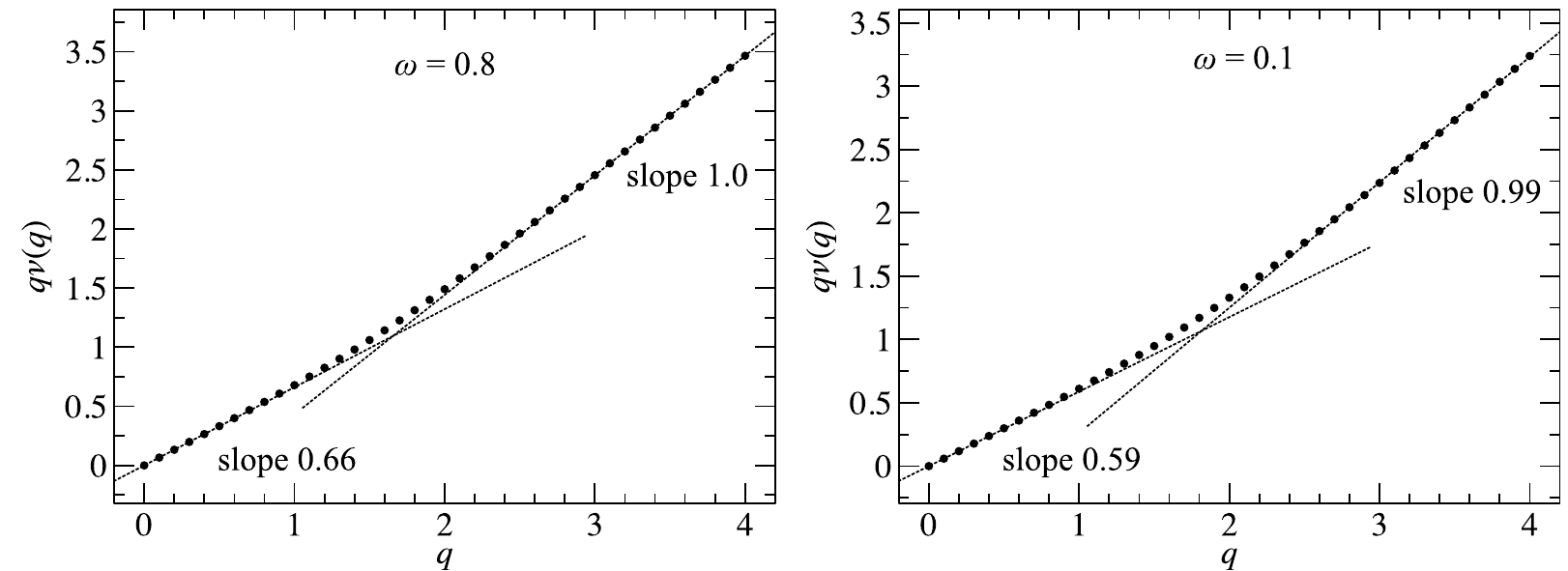}
\caption{Asymptotic exponent of the fractional moments as defined by eq.~(\ref{eq:qnu}). Strong anomalous diffusion is made evident for both values of $\omega$ by the existence of two regimes: superdiffusive and ballistic for lower and higher values of $q$, respectively. Normal diffusion would correspond to the straight line through the origin with slope $1/2$.}
\label{fig:qnu}
\end{center}
\end{figure*}

With this caveat in mind, we select the largest time window permitted by our data, $\tau\in[10^2,10^5]$, and use that region to evaluate $q\nu(q)$. The results for our chaotic pendulum are reported in fig.~\ref{fig:qnu}. If we fit a power law to $m_q(\tau)$ over another time interval, the precise values of the exponents change, but the fact that oscillations for different $q$'s are not in phase tells us that for all but a carefully picked set of choices we find the same qualitative behaviour, where we easily recognise the hallmarks of strong anomalous diffusion: moments with low index $q$ are dominated by the most likely trajectories and exhibit superdiffusion due to the aforementioned mixed nature of the phase space; high-$q$ moments, on the other hand, are dominated by the ballistic behaviour of the extremal traces, those that have spent most time around the regular regions and have therefore traveled furthest (a detailed account of the behaviour of the solutions $\theta(\tau)$ is given in the next section). 

\begin{figure}
\onefigure{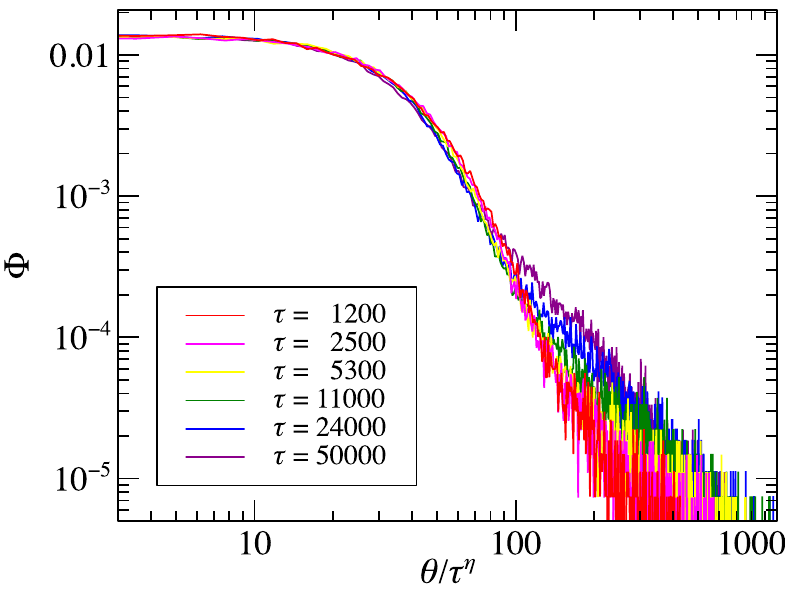}
\caption{Rescaled probability distribution $\Phi(\theta/\tau^\eta)$ for $\omega=0.8$. The value $\eta=0.61$ was found by empirically optimising the collapse and is quite close to $\lim_{q\to0}\nu(q)\approx 0.66$.}
\label{fig:coll}
\end{figure}

Another way to visualise the nature of the diffusion is to consider the probability density function $P(\theta,\tau)$ for the phase $\theta$ at time $\tau$. In the case of a normal diffusion process ($\nu(q)=1/2$ for any $q$), we would expect to be able to find an exponent $\eta$ and a function $\Phi$ such that 
\begin{equation} \label{eq:scal}
	P(\theta,\tau) = \frac{1}{\tau^\eta}\Phi\left(\frac{\theta}{\tau^\eta}\right)
\end{equation} 
with $\eta=1/2$. For weak anomalous diffusion ($\nu(q)=\nu$ constant but different from $1/2$) the scaling form above still holds, but in general $\eta\neq\nu$. In the case of strong anomalous diffusion ($\nu(q)$ not constant), however, it is not possible to find $\eta$ such that eq.~(\ref{eq:scal}) is satisfied for all the values of $\theta$: the collapse of the curves $P(\theta,\tau)$ for different $\tau$ is limited to the central part of the distribution and breaks down in the large-$\theta$ regime (see fig.~\ref{fig:coll}) dominated by ballistic events for which the scaling variable would rather be $\theta/\tau$.   

In ref.~\cite{Schmiedeberg:2009} it is showed that for a class of continuous-time random walks it is $\eta=\lim_{q\to 0}\nu(q)$. Here we compare the slope 0.66 from fig.~\ref{fig:qnu} with $\eta=0.61$ which gives the best superposition of the curves $\Phi(\theta/\tau^\eta)$ for $\omega=0.8$. In the case $\omega=0.1$ the same comparison can be drawn between 0.59 from the right panel of fig.~\ref{fig:qnu} and the empirically determined $\eta=0.54$.

\begin{figure*}
\begin{center}
\includegraphics{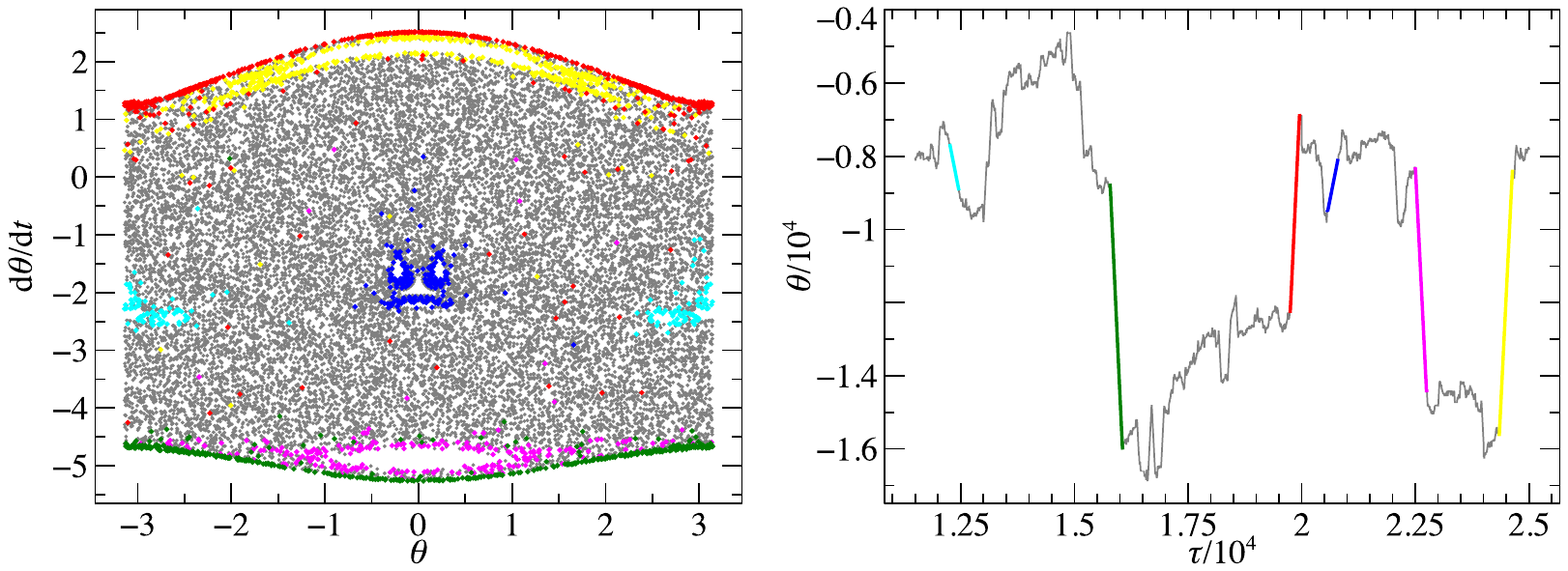}
\caption{On the left panel, the Poincar{\'e} section of one trajectory, obtained with $\omega=0.8$; on the right panel, a segment of the same trajectory. The colours highlight regions of the phase space where the dynamics gets trapped, so that the velocity stays approximately constant for a macroscopically long time.}
\label{fig:tr}
\end{center}
\end{figure*}

\section{Phase space and trajectories}
In the previous comments about the diffusion properties we have repeatedly emphasised the role played by the phase space $(\theta,\dot{\theta})$; let us now try to visualise its structure and how such structure is reflected by the trajectories $\theta(\tau)$. 
In the left panel of fig.~\ref{fig:tr} a solution of eq.~(\ref{eq:pend}) obtained for $\omega=0.8$ and initial condition close to $(0,0)$ is represented by its Poincar\'e section: the phase $\theta$ and angular velocity $\dot{\theta}$ are recorded at the end of every cycle of the external force. On the right panel of the same figure we draw a segment of the $\theta(\tau)$ graph of the same solution. We clearly recognise the typical features of a mixed phase space: even though regular regions are inaccessible to a chaotic trajectory (and are therefore indicated by blank spaces in the Poincar\'e section), their presence influence the dynamics by creating traps whence, due to barriers formed by cantori, it takes a long time for the system to get out. While trapped in these regions the pendulum maintains a constant velocity and as a consequence large variations of $\theta$ occur, in contrast with the random walk that takes place while the system is exploring the so called chaotic sea.

\begin{figure}
\onefigure{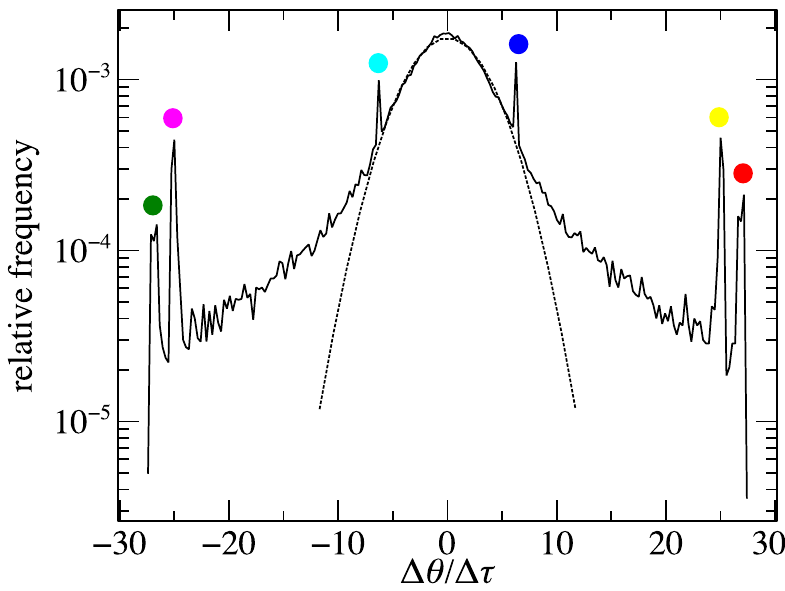}
\caption{Distribution of $\Delta\theta/\Delta\tau$ obtained for $\omega=0.8$ with $\Delta\tau=50$. The dashed line is a Gaussian approximation to the central part. The typical velocities of the six coloured regions in fig.~\ref{fig:tr} are signaled by the peaks: from left to right $B_3^-$ (green), $B_2^-$ (magenta), $B_1^-$ (cyan), $B_1^+$ (blue), $B_2^+$ (yellow) and $B_3^+$ (red).}
\label{fig:histVel}
\end{figure}

The deterministic trajectory in fig.~\ref{fig:tr}, right panel, resembles the result of a stochastic process belonging to the family of L\'evy walks~\cite{Zaburdaev:2015}. Following the method of ref.~\cite{Denisov:2002}, we can explicitly construct such process. As a first step, we need to identify a small number of sticky regions, which is easily accomplished by considering the distribution of $\Delta \theta=\theta(\tau+\Delta\tau)-\theta(\tau)$ for a suitable fixed value of $\Delta\tau$. In the case $\omega =0.8$ the choice $\Delta \tau =50$ permits the easy identification of six peaks symmetrically arranged around 0  (see fig.~\ref{fig:histVel}), which match the six coloured regions in fig.~\ref{fig:tr}. We can then perceive the excursions in the chaotic sea, during which $\theta$ undergoes normal diffusion with zero mean, as pauses between ballistic flights, which happen when the system is trapped
 into one of the six regions $B_i^\pm$ ($i=1,2,3$). See the caption of fig.~\ref{fig:histVel} for the correspondence between colour codes and typical mean velocities $\Delta\theta/\Delta\tau$ (not to be confused with the instantaneous velocities $\dot{\theta}\equiv\rmd\theta/\rmd t$). 

The time between flights is exponentially distributed $\psi\sub{c}(t)=\tau\sub{c}^{-1} \rme^{-t/\tau\sub{c}}$ (in our case $\tau\sub{c}\approx 270$ cycles of the external force), while the duration of each flight follows a power-law $\psi_i(t)\sim t^{-\alpha_i-1}$. To each ballistic region we can therefore associate three quantities: 1) a winding number $w_i\equiv\Delta\theta/(2\pi\Delta\tau)$, which is the velocity maintained during the flight, 2) the probability $p_i$ of entering region $i$ from the chaotic sea and 3) the exponent $\alpha_i$ that characterises the residence time within the region. We have already identified the winding numbers as the peaks in fig.~\ref{fig:histVel}; as for the probabilities $p_i$ and residence time exponents $\alpha_i$, we divide the time in intervals of duration $\Delta\tau=10$ cycles of the external force and associate each interval to one ballistic region or to the chaotic sea according to the mean velocity falling into one of the peaks or not.  The parameters that better describe the phase diffusion of our chaotic pendulum for $\omega=0.8$ are summarised in table~\ref{tab:km}. Observe that for the regions $B_2^\pm$ we report two exponents, which dominate the behaviour of $\psi_2(t)$ for $t< 80$ ($\alpha_2=0.5$) and for $t>80$ ($\alpha'_2=2.5$). Of course, the identification of just six ballistic regions is a simplification of the real complexity of the phase space structure: it may well be that the velocity-based approach we are adopting here simply does not have enough resolving power to distinguish regions with similar winding number but quite different residence time distributions, and this is reflected in the lack of a definite power law for regions $B_2^\pm$. 

\begin{table}
\caption{Parameters of the kinetic model that mimics the diffusion properties of the phase of the chaotic pendulum for $\omega=0.8$.
The system diffuses during an exponentially distributed time, and then enters with probability $p_i$ one of the six ballistic regions characterised by winding number $w_i$ and residence time exponent $\alpha_i$.
}
\label{tab:km}
\begin{center}
\begin{tabular}{lccc}
 & $w_i$ & $p_i$ & $\alpha_i$ \\
\hline \\
$B_1^\pm$ & $\pm1.0$ & 0.296 & 1.5 \\
$B_2^\pm$ & $\pm4.0$ & 0.197 & 0.5 / 2.5 \\
$B_3^\pm$ & $\pm4.2$ & 0.007 & 0.5
\end{tabular}
\end{center}
\end{table}

Note that previous studies of the phase space structure of this same system~\cite{Sakthivel:2011} focused on the extreme velocity regions $B_2^\pm$ and $B_3^\pm$, overlooking the role played by $B_1^\pm$ which we on the contrary find determinant to describe the superdiffusive regime. Actually, from the data in table \ref{tab:km} we can estimate a theoretical value for the diffusion exponent $2\nu(2)$. As the presence of exponentially distributed waiting times between ballistic flights has no effect on the mean squared displacement~\cite{Cristadoro:2014}, we can use the scaling laws~\cite{Geisel:1985,Shlesinger:1985}  
\begin{equation} \label{eq:cases}
\langle\theta(t)^2\rangle \sim \begin{cases}
	t^2 & 0<\alpha<1 \\
	t^2/\ln t & \alpha = 1 \\
	t^{3-\alpha} & 1 < \alpha < 2 \\
	t \ln t & \alpha = 2 \\
	t & \alpha > 2
\end{cases}
\end{equation}
derived for a L\'evy walk characterised by the flight time distribution $\psi(t)\sim t^{-\alpha-1}$. 

The probability distribution of the residence time in $B_2^\pm$ falls off rapidly (exponent 2.5) for large $t$, so we can assume these regions not to play a dominant role in determining the diffusion exponent. On the other hand, even if regions $B_3^\pm$ host the longest ballistic excursions, they are rarely accessed: we expect them to influence the extreme trajectories (relevant to moments of larger order), but the mean squared displacement should be dominated by the more frequently visited $B_1^\pm$ regions. If we adopt $\alpha_1=1.5$ as the exponent that better characterises the phase diffusion process, then eq.~(\ref{eq:cases}) yields $\langle\theta(t)^2\rangle\sim t^{1.5}$, which is in fact the value we observe (see fig.~\ref{fig:qnu}). 

The same analysis can be applied to the case $\omega=0.1$ with similar results: the main difference is that the kinetic model is simpler as we need to identify only two ballistic regions $B^\pm$, characterised by winding number $w=\pm 140$  rounds per cycle of the external force and by the exponent $\alpha=1.7$. Once again the expectation for the diffusion exponent based on eq.~(\ref{eq:cases}) matches the value 1.3 from our numerical test.

Another quantity that can be analytically computed in a L\'evy walk model is the behaviour of $\nu(q)$ for small $q$, for which we have the relation~\cite{Andersen:2000} $\lim_{q\to 0}\nu(q)=1/\alpha$. Taking $\alpha_1=1.5$ as the dominant exponent for $\omega=0.8$ and $\alpha = 1.7$ for $\omega=0.1$ we would expect therefore that $\nu(q)=0.67$ and $0.59$, respectively, when $q$ is small, in remarkable agreement with fig.~\ref{fig:qnu}. Note that the slope for small $q$ in the figure is not calculated from a fit, whose result would depend on how many points are taken into account, but from tracing the straight line between the first two data available: $q=0$ and $q=0.1$. This gives another check that the kinetic model we have built is consistent with the results of our numerical analysis.


\section{Conclusions} 
We have presented numerical evidence that the phase of the harmonically driven undamped pendulum of eq.~(\ref{eq:pend}) undergoes strong anomalous diffusion in both the cases we have analysed ($\gamma =1.2$, $\omega=0.1$ or 0.8). We have attributed such phenomenon to the mixed nature of the phase space, with sticky regions near the regular solutions that act as dynamical traps, so that the time evolution of $\theta$ resembles that of a stochastic process where standard diffusion is interrupted by ballistic flights with power-law distributed duration. This slow decay of the flight-time distribution is responsible for the superdiffusive behaviour, while the existence of extremal flight velocities determines the strongly anomalous character of the diffusion process, as can be seen in analytical calculations based on a L\'evy walk model~\cite{Rebenshtok:2014,Zaburdaev:2015}.

With the idea of suggesting a system where strong anomalous diffusion could be experimentally observed, we plan in the future to investigate to what extent these features survive the introduction of a friction term.   

\acknowledgments
Part of this research took place at the Galileo Galilei Institute for Theoretical Physics in Arcetri, during the INFN-funded summer 2014 workshop \emph{Advances in Nonequilibrium Statistical Mechanics}.

\end{document}